\documentclass[a4paper,fleqn,usenatbib]{mnras} 

\pdfoutput=1
\usepackage[T1]{fontenc}
\usepackage{ae,aecompl}
\usepackage[export]{adjustbox}
\usepackage{graphicx}
\usepackage{times}

\usepackage[pscoord]{eso-pic}
\usepackage{amsmath}
\usepackage{amssymb}

\def\gsim{ \lower .75ex \hbox{$\sim$} \llap{\raise .27ex \hbox{$>$}} }
\def\lsim{ \lower .75ex \hbox{$\sim$} \llap{\raise .27ex \hbox{$<$}} }

 \title[DM Review]{Cosmological Dark Matter: a Review (the April Fool Edition)}
\author[M.~R.~Lovell et al.]{Mark R. Lovell\thanks{email: lovell@hi.is}$^{1}$\\
$^{1}$Center for Astrophysics and Cosmology, Science Institute, University of Iceland, Dunhagi 5, 107 Reykjavik, Iceland}

\date{Accepted ... Received ...; in original form ...} 
\pubyear{2020}

\begin{document}

\label{firstpage}
\pagerange{\pageref{firstpage}--\pageref{lastpage}} 
  
\maketitle

\begin{abstract}

\noindent Evidence has continued to accumulate over the last few decades as to the existence and nature of dark matter. Depending on the particle candidate, the dark matter can exhibit one of several cosmologically defined models: hot dark matter, cold dark matter, warm dark matter, self-interacting dark matter, and fuzzy dark matter. In this paper I review the relevance and status of these models, whether it is possible for more than one of these models to each constitute some fraction of the dark matter, and discuss the prospects for determining if any of these models can successfully describe the properties and evolution of  our own Universe. 

\end{abstract}


\section{Introduction}
\label{intro}

The first evidence for the existence of matter that does not interact with light was presented in observations of the Coma cluster in the 1930s by Fritz Zwicky, who measured the velocities of the Coma member galaxies and determined that they were moving faster than their combined stellar mass could bind together gravitationally. Further evidence was announced in the 1970s with Vera Rubin's measurements of galaxy rotation curves, which showed that the total mass of galaxies increased significantly with radius even though very little additional stellar mass was present at these larger radii. The inference of large amounts of non-gravitating mass from gravitational lensing in clusters, plus the statistics of the CMB, has left the existence of dark matter as a key component of the matter-energy density of the Universe. 

This subject can be approached from the view of either particle physics or of cosmology. First, in the particle physics case we are concerned with the nature of the dark matter candidate and its interactions with the standard model of particle physics. Candidate particles have included supersymmetric neutralinos, gravitinos, QCD axions, axion-like particles (ALPs), sterile neutrinos, standard model neutrinos, and even small black holes. The second approach is instead to consider cosmological definitions of dark matter based on how they influence the seeding and evolution of galaxies. Candidates models under thus framework include hot dark matter (HDM), cold dark matter (CDM), warm dark matter (WDM), self-interacting dark matter (SIDM) and fuzzy dark matter (FDM). 

In this study we take the second approach, considering the implications of dark matter for the formation of structure. From the cosmological point of view, it remains very unclear which of the dark matter models is the best. There is a large degree of degeneracy between dark matter physics and baryonic processes such as feedback from supernovae and active galactic nuclei, the heating of gas in small haloes during reionization, and even cosmic rays and magnetic fields. Part of the difficulty of analysing such models is that the behaviour of even the simplest dark matter models is quite unlike that of other materials on planet Earth. Thankfully, of all the places that do exist on Earth the island of Iceland is that perhaps most akin in substance and character to some location from another planet\footnote{But not from alternative fantasy settings such as Middle Earth or Westeros: due to the introduction of MNRAS page charges this article is specifically limited to sci-fi.} and is therefore an ideal place to undertake dark matter studies. In this article I explain the current status, positives and negatives of each of these five models models -- HDM, CDM, WDM, SIDM, FDM -- and later draw conclusions. 
 
\section{Review} 
\label{rev}

\subsection{Hot dark matter}
\label{sims:hdm}

\begin{figure}
	\includegraphics[scale=0.065]{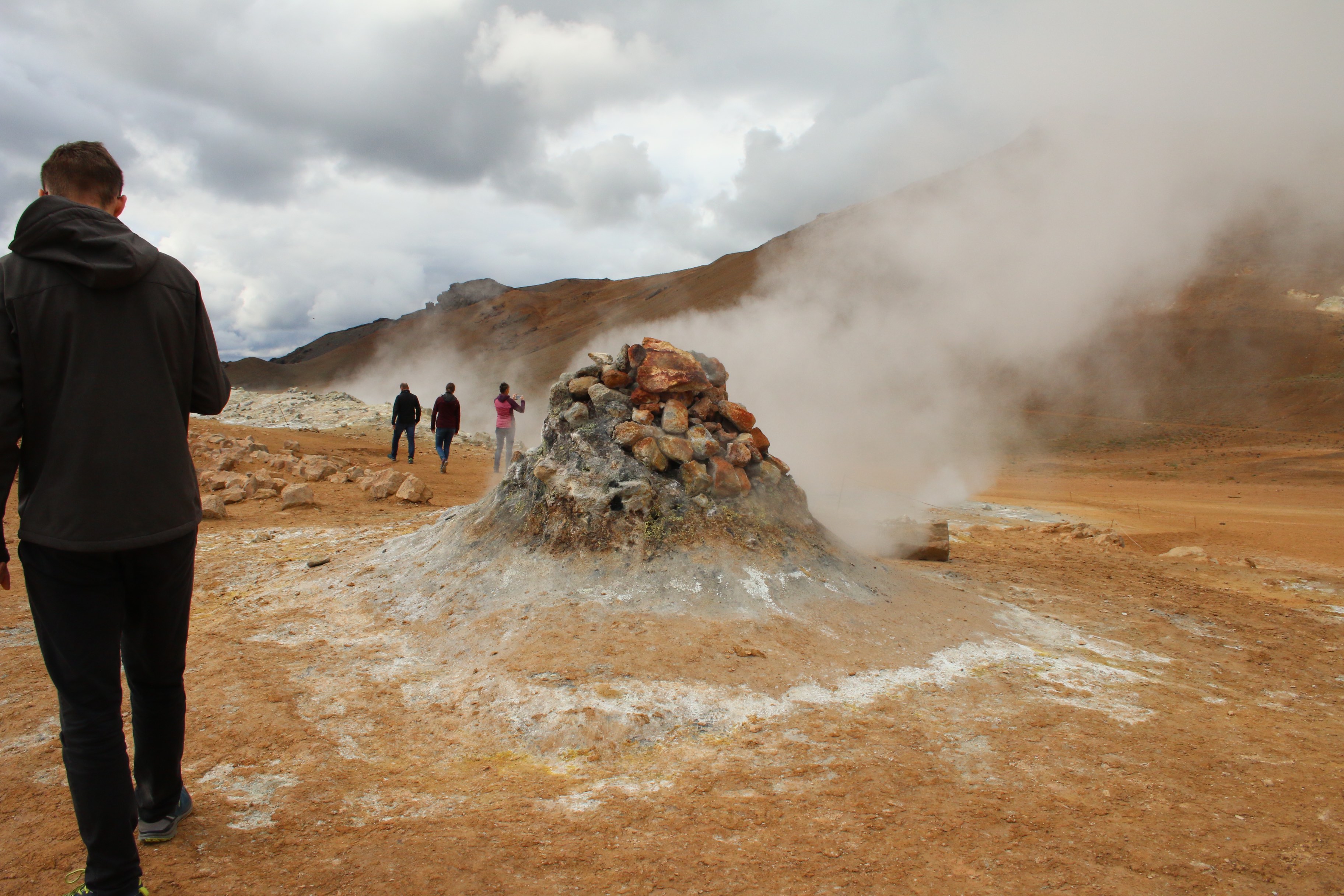}
	 \caption{Image of the emission of HDM. HDM is disfavoured, largely because it smells awful.}
	 \label{fig:hdm}
 \end{figure}

HDM was first proposed in the early 1980s, largely by Soviet scientists. An image is included in Fig.~\ref{fig:hdm}. In HDM the first structures to form are galaxy clusters, which then fragment into smaller galaxies over time. In later years, although still a long time ago (1985), it was shown that the true distribution of galaxies is much less clustered than HDM simulations predicted, therefore the HDM model was discontinued at that time, together with the Lada 1500. The reader should be warned that some very small proportion of the dark matter is in fact known to be must be HDM, since standard neutrinos have been shown to have mass. Given HDM's contacts to the former Soviet Union, it goes without saying that, if offered HDM on the street, just say no, and immediately inform the police. 

\subsection{Cold dark matter}
\label{sims:cdm}

\begin{figure}
	\includegraphics[scale=0.065]{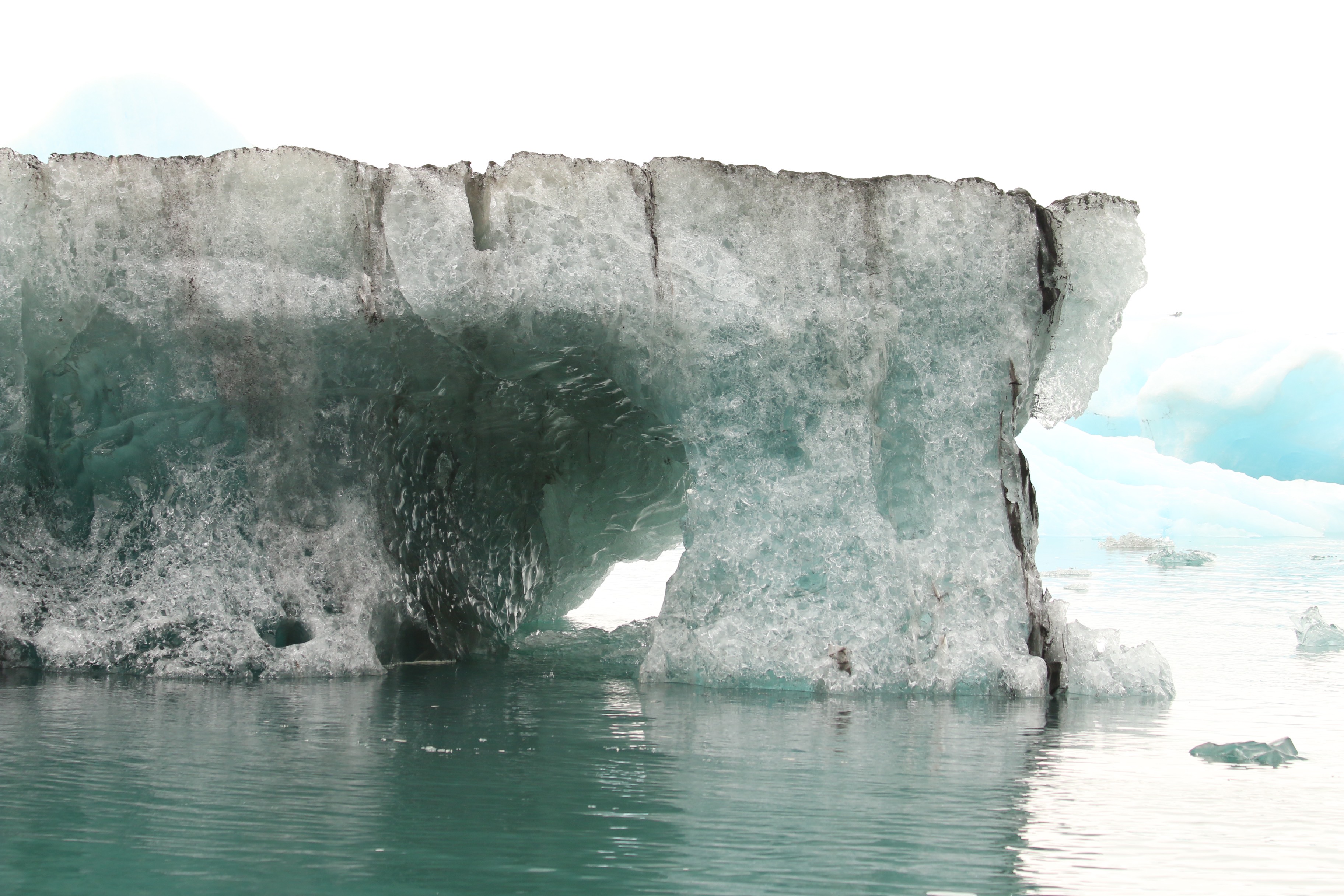}
	 \caption{A comparison between idealised CDM (top of image) and the CDM model with first order fluid corrections (bottom of image).}
	 \label{fig:cdm}
 \end{figure}

CDM has been a popular dark matter model for much of the last forty years, especially since HDM fell out of favour. It assumes that the dark matter has negligible thermal velocity, and in its idealised form is considered to have no thermal velocity at all, which makes it very easy to simulate with $N$-body codes. In practice, realistic CDM models in our Universe will inevitably exhibit some very small velocity, and so behave like a fluid that is not truly, completely cold. We illustrate this difference between realistic and idealised CDM models in Fig.~\ref{fig:cdm}. 

 This model has done a much better job than HDM at describing our Universe, including the properties and distributions of galaxies, the statistics of the cosmic microwave background, and the lensing patterns around galaxy clusters. However, the failure to find this matter in experiments, whether deep underground, in particle accelerators, or in attempts to detect gamma ray annihilation emission from Galactic or extragalactic targets. This study suggests that alternative equipment may be required to study CDM. Like a boat. To conclude, despite early promise, CDM is not the messiah, nor even a naughty boy: it might not even exist...  

\subsection{Warm dark matter}
\label{sims:wdm}

\begin{figure*}
	\includegraphics[scale=0.055]{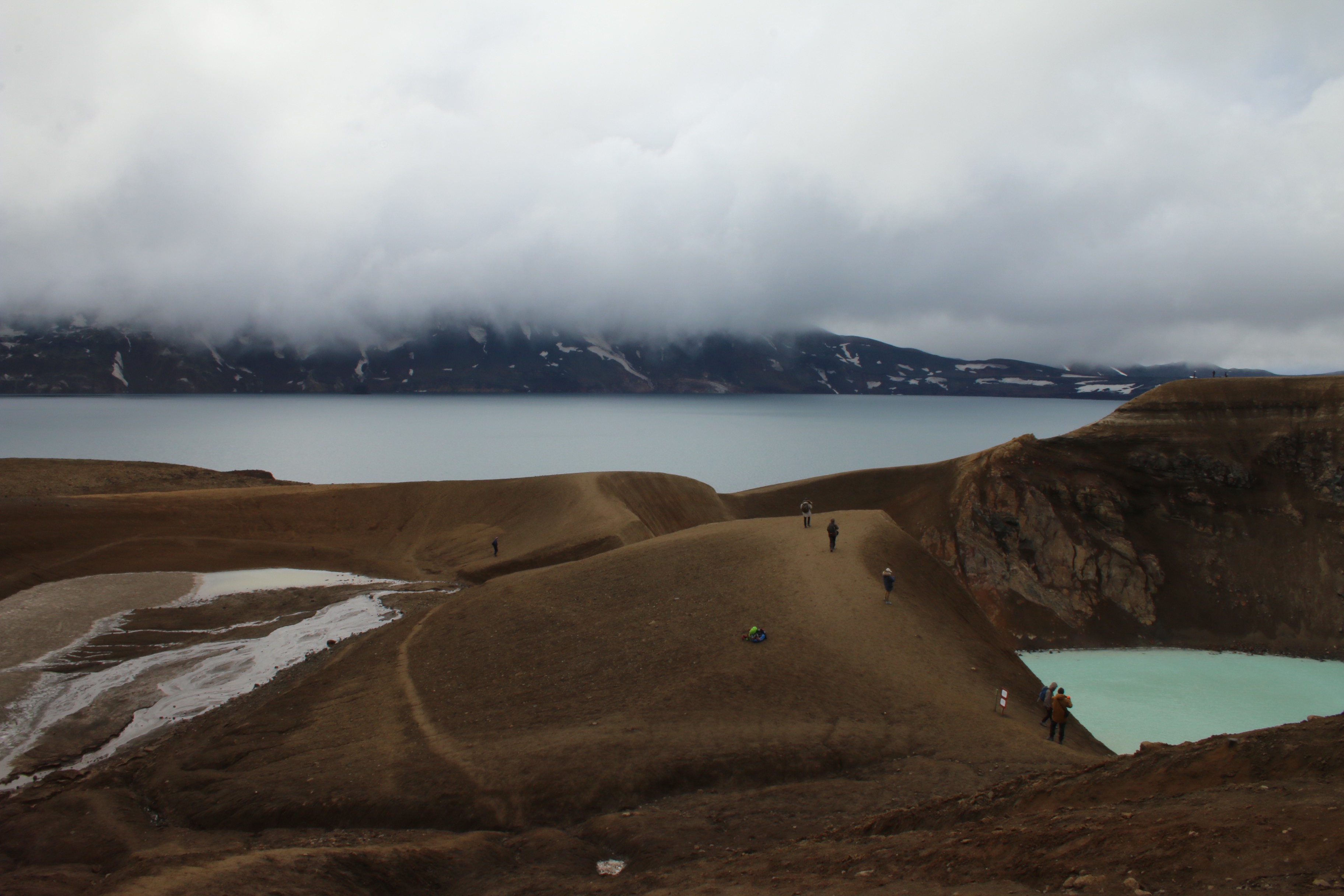}
	\includegraphics[scale=0.055]{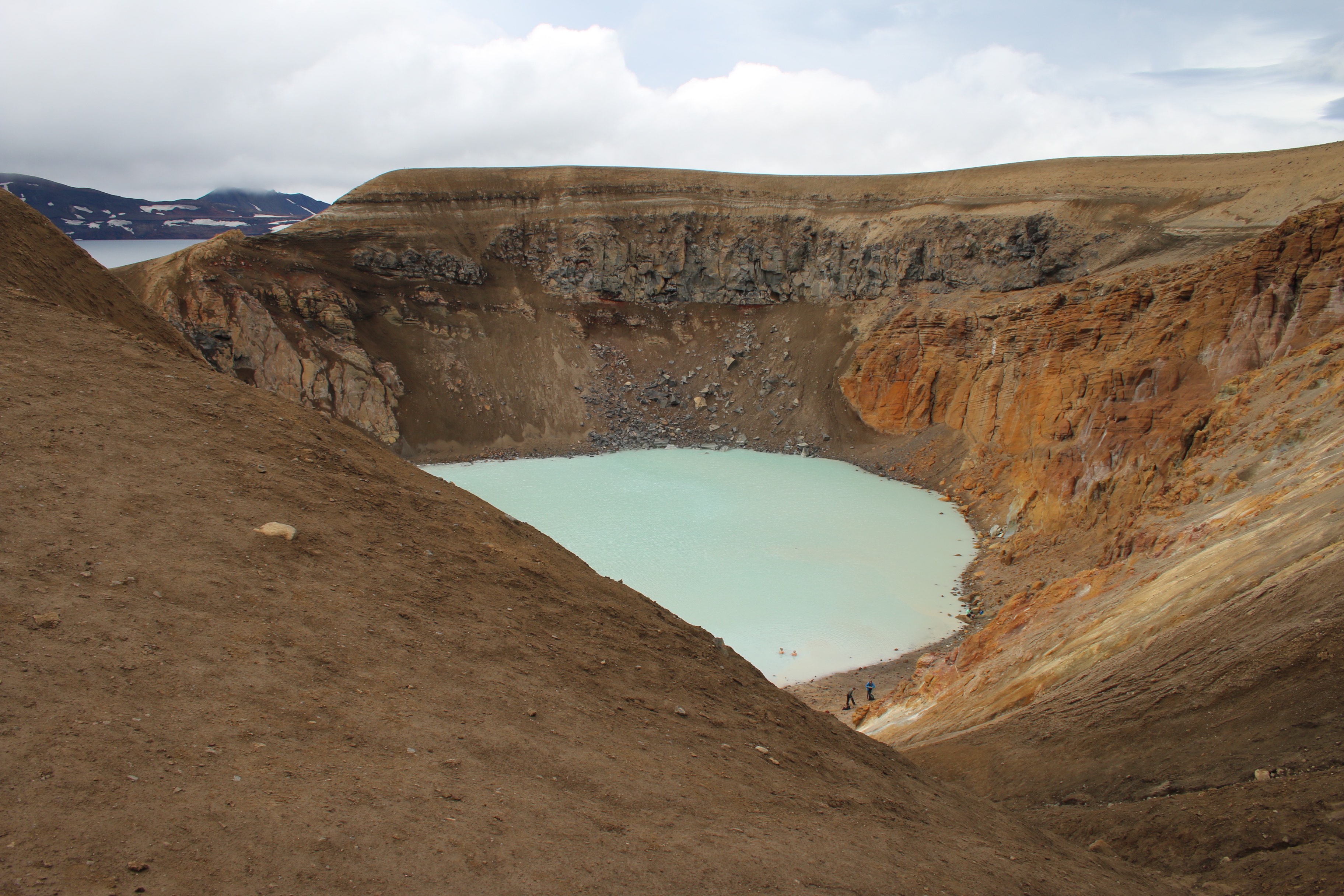}
	 \caption{In the left-hand panel, we compare CDM (including fluid corrections, image centre) with WDM (bottom right of image).  WDM is subsequently found to be the superior model, because, as we show in the right-hand panel, it is possible to swim in it.}
	 \label{fig:wdm}
 \end{figure*}

WDM is a most impressive dark matter model. It has grown increasingly popular in recent years, since with the passage of time we put space between us and the Iraq war, and thus from the suggestion that WDM is a misspelling of WMD.  It shares many of its properties with CDM, but has the crucial difference that its thermal velocities are explicitly much greater than zero and can impact galaxy formation. It sits on a spectrum between HDM and the realistic CDM models, which we illustrate in Fig.~\ref{fig:wdm}. Those extra thermal velocities have the effect of suppressing the number and density of low mass haloes and forces them to form later.  Also, WDM models can be part of larger theories that also predict the overabundance of matter over antimatter in the Universe and also explain why neutrinos -- the HDM neutrinos described above -- have mass\footnote{We do not invoke the extra benefit that decaying WDM can emit detectable X-ray signals in this paper, because the 3.55~keV line is not an April Fools joke.}. There are remarkable hints that WDM models may provide a better match to the properties of observed galaxies than CDM, such as the densities of Local Group dwarf galaxies, but the arguments over whether it is really baryonic feedback and non-detection of faint CDM dwarfs will likely rage for some while longer. Although we have discussed here how WDM is generally fantastic, nutritious and wholesome, and an excellent dark matter model, it must be noted that it can be somewhat addictive. Evidence for the prevalence of WDM dependence can be found in Lovell~et~al. (2012, 2014, 2015, 2016, 2017a, 2017b, 2019, 2020a, 2020b, 2020c...)

\subsection{Self-interacting dark matter}
\label{sims:sidm}

\begin{figure}
	\includegraphics[scale=0.2]{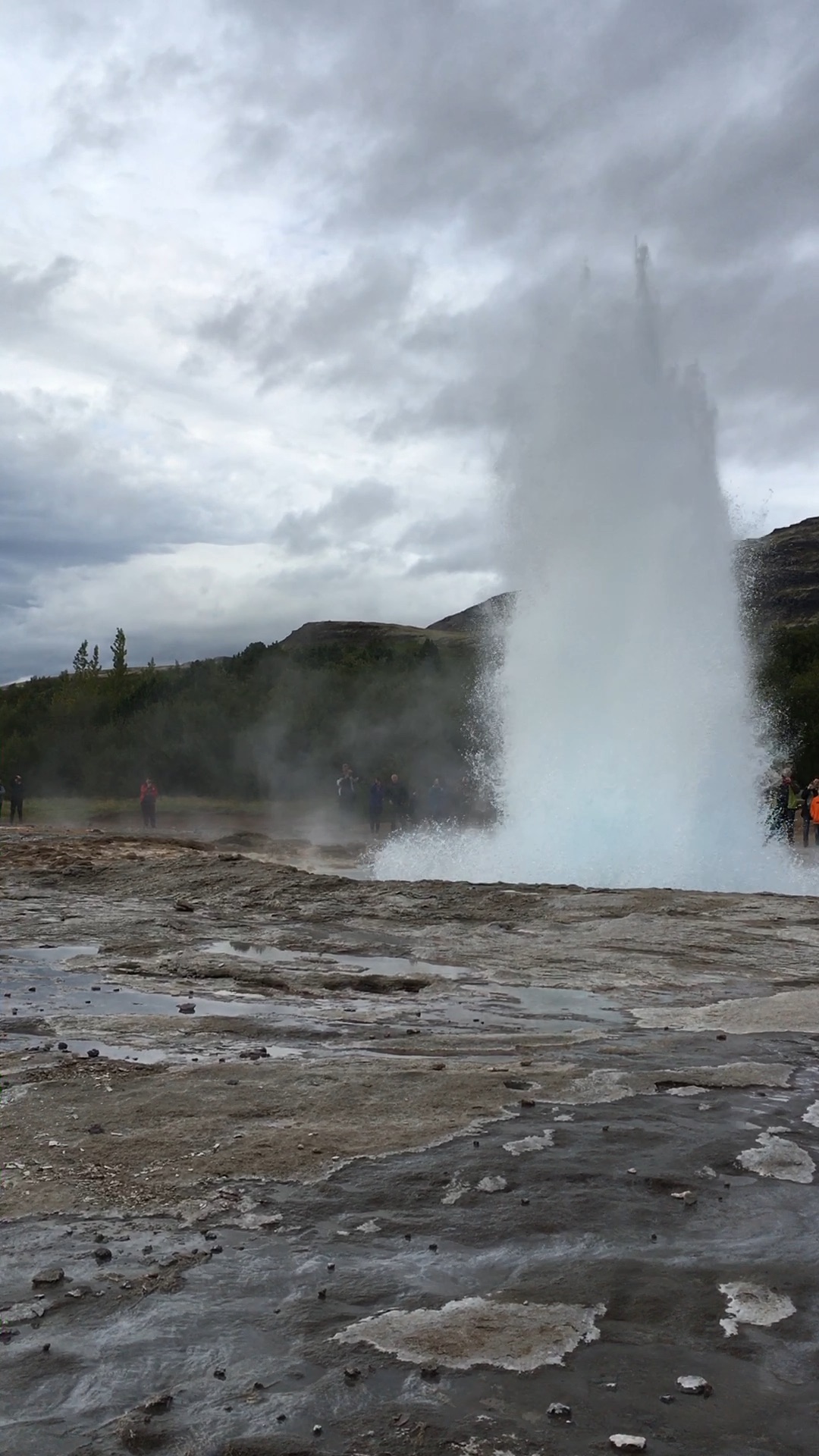}
	 \caption{CDM exhibits interactions of the same magnitude as the strong nuclear force, which we show here. It remains to be seen whether self-interactions effect such bursty core formation as is the case for star-formation.}
	 \label{fig:sidm}
 \end{figure}

SIDM has been envisaged as a model that serves primarily to explain why some galaxies appear to have a dark matter core. Whether or not such cores exist is debated, and debated, and debated, and debated some more, so here we will put it down as a `maybe'. It is envisaged under this scenario that the dark matter particles undergo self-interactions, and can scatter off one another strongly. We present an example of this sort of interaction in Fig.~\ref{fig:sidm}. Such self-interactions can even evaporate satellite galaxies if too strong, therefore handling SIDM and its parameters requires great care. Possibly with gloves. One of the motivations behind adopting a model with self-interactions is the suggestion that the dark matter may comprise of many particles that constitute a whole `dark sector', featuring dark atoms and dark radiation. Given the difficulty in identifying even one candidate dark matter particle, let alone a whole mirror standard model's worth, it therefore appears that most SIDM-studying scientists are masochistic. This statement is true provided that reality as a whole is not barred from being sadistic. In which case the SIDM enthusiasts would just be Right. And would be able to keep particle physicists occupied for decades.

\subsection{Fuzzy dark matter}
\label{sims:hdm}

One of the primary issues facing dark matter studies is that the standard model of particle physics has very few particles to choose from that could constitute dark matter. Some more expansive models would, however, inevitably lead to many more candidates, so the key was to find a particle physics model that paints a picture big enough to contain a suitable candidate. It was inevitable that once physicists had brought home the string theory landscape and unrolled it on their desks that a dark matter candidate would fall out and onto the floor.  These dark matter candidates were expected to be very light, and to show quantum mechanical properties at macroscopic scales, even on the scales of dwarf galaxies. The dark matter will exhibit macroscopic wavefunctions with peaks and troughs.
By $z=0$, these quantum interactions lead to the generation of cores, and therefore in some ways FDM is  in competition with SIDM. The cores are surrounded by very steep density profiles which is somewhat unusual. These very steep cores are remarkably similar to what happens with the contraction and expansion of baryons. This is beneficial, because if we can show that FDM matter is actually baryons then this would save us all a lot of time; although that would then mean that we would all have to work on dark energy instead and that topic looks rather difficult. 


\section{Summary \& conclusions}
 \label{conc}
 
It is a truth universally acknowledged that a physicist in possession of a grounding in particle physics and astronomy be in want of the one true dark matter particle identity. In this paper we have reviewed the properties of five different dark matter models: HDM, CDM, WDM, SIDM and FDM. We showed how HDM was initially very popular but has encountered insurmountable difficulties to be considered as 100~per~cent of the dark matter. CDM has shown much more promise, including the ability to match the distribution of observed galaxies, and in principle is almost as good as WDM.  We discussed how SIDM and FDM both generate cores, albeit in very different ways. They are therefore potentially more exciting than the observed Universe, which still might not make cores: am holding fast too that `maybe' cited in the previous section.  We conclude that many of these modules still have a future. Just not a bright future. Because they are dark matter.

\section*{Acknowledgements}

 MRL would like to thank the late Terrys Pratchett and Jones. MRL would like to apologise to Jane Austen. The Figure locations are as follows: Fig.~\ref{fig:hdm}, Hverir, North Iceland; Fig~\ref{fig:cdm}, J\"okuls\'arl\'on,  South Iceland; Fig~\ref{fig:wdm}, Askja caldera lake ft. V\'iti crater, Highlands of Iceland; Fig.~\ref{fig:sidm}, Strokkur geyser, South Iceland. 

\bibliographystyle{mnras}
\bibliography{bibtex}
    
\bsp
\label{lastpage}

\end{document}